\documentclass{epsconf}

\usepackage{amsmath}
\usepackage{amssymb}
\usepackage{stmaryrd}

\def\wt#1{\widetilde{#1}}
\def\vb#1{\mbox{\boldmath$#1$}}
\def\pd#1#2{\frac{\partial #1}{\partial #2}}

\def\wh#1{\widehat{#1}}
\def\bdot{\,\vb{\cdot}\,}
\def\btimes{\,\vb{\times}\,}

\def\bhat{\wh{{\sf b}}}
\def\cal#1{\mathcal{#1}}

\def\bhat{\wh{{\sf b}}}
\def\exd{{\sf d}}

\newcommand{\bc}{\begin{center}}
\newcommand{\ec}{\end{center}}
\newcommand{\bt}{\begin{tabbing}}
\newcommand{\et}{\end{tabbing}}
\newcommand{\be}{\begin{eqnarray*}}
\newcommand{\ee}{\end{eqnarray*}}
\newcommand{\bs}{\begin{slide}}
\newcommand{\es}{\end{slide}}

\title{Exact conservation laws for \\ truncated gyrokinetic Vlasov-Poisson equations}

\author{Natalia Tronko$^{1}$ and Alain J.~Brizard$^{2}$}
\institute{$^{1}$York Plasma Institute, University of York, Heslington, York, YO10 5DD, UK  \\ $^{2}$Department of Physics, Saint Michael's College, Colchester, VT 05439, USA}

\begin{document}
\maketitle

Exact conservation laws for the gyrokinetic Vlasov-Poisson equations can either be derived from a variational principle by the Noether method \cite{Brizard_Tronko} or directly if exact invariants for gyrocenter Hamiltonian dynamics are known \cite{Scott_Smirnov}. We begin our Noether derivation with the noncanonical gyrocenter phase-space Lagrangian
\begin{equation}
\Gamma_{\rm gy} \;\equiv\; \left[\left( \frac{e}{c}\,{\bf A} \;+\; p_{\|}\,\bhat \right)\bdot\exd{\bf X} \;-\; W\;\exd t\right] \;-\; \left(
H_{\rm gy} - W\right)\; \exd\tau,
\label{eq:Gamma_gy}
\end{equation}
where $W$ is the extended phase-space energy coordinate and the gyrocenter Hamiltonian is
\begin{equation}
H_{\rm gy}({\bf X}, p_{\|}, \mu, t;\; \Phi_{1}) \;\equiv\; \mu\,B + \frac{p_{\|}^{2}}{2m} \;+\; \epsilon\,e\;\langle\Phi_{1{\rm gc}}\rangle.
\label{eq:H_gy}
\end{equation}
The gyrocenter Euler-Lagrange equation associated with an arbitrary displacement $\delta{\bf X}$ is
\begin{equation}
\frac{e}{c}\;\frac{d_{\rm gy}{\bf X}}{dt}\btimes{\bf B}^{*} \;-\; \frac{d_{\rm gy}p_{\|}}{dt}\;\bhat \;-\; \nabla H_{\rm gy} \;=\; 0,
\label{eq:X_EL}
\end{equation}
where ${\bf B}^{*} \equiv {\bf B} + (p_{\|}c/e)\;\nabla\btimes\bhat$ and the gyrocenter canonical momentum is ${\bf p}_{\rm gy} \equiv (e/c)\,{\bf A}
+ p_{\|}\,\bhat \equiv (e/c)\,{\bf A}^{*}$, from which we obtain the Hamilton equation for the canonical momentum \cite{Brizard_Tronko}
\begin{equation}
\frac{d_{\rm gy}{\bf p}_{\rm gy}}{dt} \;=\; -\;\nabla\,H_{\rm gy} \;+\; \frac{e}{c}\,\nabla{\bf A}^{*}\bdot\frac{d_{\rm gy}{\bf X}}{dt}.
\label{eq:pgy_dot}
\end{equation}
In axisymmetric tokamak geometry, the magnetic vector potential is ${\bf A} \equiv -\,\psi\,\nabla\varphi + \Psi(\psi)\,\nabla\vartheta$, so that the magnetic field ${\bf B} \equiv \nabla\varphi\btimes\nabla\psi + q(\psi)\;\nabla\psi\btimes\nabla\vartheta$ yields the identity $\nabla\psi \equiv
{\bf B}\btimes\partial{\bf X}/\partial\varphi$.

We now derive the Hamilton equation for the toroidal canonical gyrocenter momentum
\begin{equation}
p_{{\rm gy}\varphi} \;\equiv\; \pd{\bf X}{\varphi}\bdot{\bf p}_{\rm gy} \;=\; -\,\frac{e}{c}\;\psi \;+\; p_{\|}\;b_{\varphi},
\label{eq:pgy_phi}
\end{equation}
where $b_{\varphi} \equiv \bhat\bdot\partial{\bf x}/\partial\varphi$ denotes the covariant toroidal component of the magnetic unit vector. By taking the toroidal projection of the gyrocenter Euler-Lagrange equation \eqref{eq:X_EL}, we obtain \cite{Brizard_Tronko} $d_{\rm gy}p_{{\rm gy}\varphi}/dt \equiv
-\,\partial H_{\rm gy}/\partial\varphi$, where we used the identity
\begin{equation}
\pd{\bf C}{\varphi} \;+\; \nabla\left(\pd{\bf x}{\varphi}\right)\bdot{\bf C} \;=\; \pd{\bf C}{\varphi} \;+\; {\bf C}\btimes\wh{\sf z} \;\equiv\; 0,
\label{eq:C_varphi}
\end{equation}
which is valid for an arbitrary vector field ${\bf C}$ in axisymmetric tokamak geometry.

The truncated gyrokinetic Vlasov-Poisson equations are derived from the action functional
\begin{equation}
{\cal A}_{\rm gy} \;=\; \int_{x^{4}}\left( \frac{\epsilon^{2}\,|{\bf E}_{1}|^{2}}{8\pi} - \frac{|{\bf B}|^{2}}{8\pi} \right) \;+\;
\frac{\epsilon^{2}}{2}\,\int_{z^{7}} F_{0}\;e\left\langle \pounds_{1}\frac{}{}\Phi_{1{\rm gc}}\right\rangle \;-\; \int_{Z^{8}}\;
{\cal F}_{\rm gy}(Z)\;{\cal H}_{\rm gy}(Z;\, \Phi_{1}),
\label{eq:Action_gy}
\end{equation}
where summation over particle species is implied, with the extended gyrocenter Hamiltonian ${\cal H}_{\rm gy} \equiv H_{\rm gy} - W$ and the extended gyrocenter Vlasov distribution ${\cal F}_{\rm gy} \equiv c\,\delta(W - H_{\rm gy})\; F$. In \eqref{eq:Action_gy}, the gyrocenter Lie derivative is 
$\pounds_{1}\Phi_{1{\rm gc}} \equiv \{ S_{1},\; \Phi_{1{\rm gc}}\}_{\rm gc}$, where $S_{1} \equiv (e/\Omega)\int\,\wt{\Phi}_{1{\rm gc}}\,d\theta$ and 
$\{\;,\;\}_{\rm gc}$ is the guiding-center Poisson bracket. The gyrokinetic variational principle $\delta{\cal A}_{\rm gy} \equiv \int\;
\delta{\cal L}_{\rm gy}\;d^{4}x = 0$ introduces the variation of the gyrokinetic Lagrangian density
\begin{equation}
\delta{\cal L}_{\rm gy} = -\epsilon\,\delta\Phi_{1} \left[ e\int_{z^{6}} \left( \langle\delta^{3}_{\rm gc}\rangle\;F - \epsilon\;F_{0}\left\langle
\pounds_{1}\frac{}{}\delta^{3}_{\rm gc}\right\rangle \right) \right]  + \frac{\epsilon^{2}}{4\pi}\;\left( \delta{\bf E}_{1}\bdot
{\bf E}_{1} \right) - \int\;\delta{\cal F}_{\rm gy}\;{\cal H}_{\rm gy}\;d^{4}p,
\label{eq:delta_L_gy}
\end{equation}
where $\delta{\cal F}_{\rm gy} \equiv \{ {\cal S}_{\rm gy},\; {\cal F}_{\rm gy}\}_{\rm gc}$ is generated by the canonical generating function ${\cal S}_{\rm gy}$ and $\delta{\bf E}_{1} \equiv -\,\nabla\delta\Phi_{1}$. From the variational principle, we obtain the gyrokinetic extended Vlasov equation $\{ {\cal F}_{\rm gy},\; {\cal H}_{\rm gy}\}_{\rm gc} = 0$, which, when integrated over $W$, yields the gyrocenter Vlasov equation \cite{Brizard_Tronko}
\begin{equation}
\pd{F}{t} \;+\; \frac{d_{\rm gy}{\bf X}}{dt}\bdot\nabla F \;+\; \frac{d_{\rm gy}p_{\|}}{dt}\;\pd{F}{p_{\|}} \;=\; 0.
\label{eq:gy_Vlasov}
\end{equation}
We also obtain the gyrokinetic Poisson equation
\begin{equation}
\frac{\epsilon\,\nabla\bdot{\bf E}_{1}}{4\pi} \;=\; e \int_{z^{6}}\left( F \langle\delta_{\rm gc}^{3}\rangle \;-\; \epsilon\,F_{0}\;
\left\langle \pounds_{1}\frac{}{}\delta_{\rm gc}^{3}\right\rangle \right) \;\equiv\; \varrho \;-\; \nabla\bdot\mathbb{P},
\label{eq:gy_Poisson}
\end{equation}
where $\varrho$ denotes the gyrocenter charge density  and the gyrokinetic polarization
\begin{equation}
\mathbb{P} \;=\; \frac{\bhat}{\Omega}\btimes\left[ \int\;F\,\left(e\;\frac{d_{\rm gy}{\bf X}}{dt} \right) d^{3}p \right] \;\equiv\; \mathbb{P}_{\rm gc}
\;+\; \mathbb{P}_{\rm gy}
\label{eq:gk_pol}
\end{equation}
includes contributions from the guiding-center polarization (from the guiding-center velocity $d_{\rm gc}{\bf X}/dt$) and the gyrocenter polatization (from the perturbed $E\times B$ velocity $\epsilon\,(c\bhat/B)\btimes\nabla\langle\Phi_{1{\rm gc}}\rangle$).

By inserting \eqref{eq:gy_Vlasov}-\eqref{eq:gy_Poisson} into \eqref{eq:delta_L_gy}, we obtain the gyrokinetic Noether equation $\delta{\cal L}_{\rm gy} \equiv \partial\Lambda/\partial t + \nabla\bdot\vb{\Gamma}$, where the Noether fields are
\begin{equation}
\Lambda \;\equiv\; \int\; {\cal S}_{\rm gy}\;{\cal F}_{\rm gy}\;d^{4}p \;\;\;{\rm and}\;\;\; \vb{\Gamma} \;\equiv\; -\; \frac{\epsilon^{2}\,
\delta\Phi_{1}}{4\pi}\;{\bf E}_{1} \;+\; \int\; \left({\cal S}_{\rm gy}\frac{}{}{\cal F}_{\rm gy}\right)\;\frac{d_{\rm gy}{\bf X}}{dt}\;d^{4}p.
\label{eq:Lambda_Gamma}
\end{equation}
The gyrokinetic Noether equation is now used to derive the gyrokinetic toroidal angular-momentum conservation law.
First, when considering arbitrary infinitesimal displacements $\delta{\bf x}$, we obtain the Noether momentum equation \cite{Brizard_Tronko}
\begin{equation}
\pd{{\bf P}}{t} \;+\; \nabla\bdot\vb{\Pi}_{\rm gy} \;=\; -\;\int F\left( \nabla H_{\rm gy} - \frac{e}{c}\,\nabla{\bf A}^{*}\bdot
\frac{d_{\rm gy}{\bf X}}{dt} \right),
\label{eq:gy_primitive}
\end{equation}
where
\begin{equation}
{\bf P} \;=\; \int\; F\,{\bf p}_{\rm gy}\;d^{3}p \;\;\;{\rm and}\;\;\; \vb{\Pi}_{\rm gy} \;=\; \int \;F\; \frac{d_{\rm gy}{\bf X}}{dt}\;{\bf p}_{\rm gy} \;d^{3}p.
\label{eq:gy_momentum_stress}
\end{equation}
We note that \eqref{eq:gy_primitive} can also be obtained as the gyrocenter-Vlasov moment of \eqref{eq:pgy_dot}. Second, we consider the infinitesimal toroidal rotation $\delta{\bf x} \equiv \delta\varphi\;\partial{\bf x}/\partial\varphi = \delta\varphi\,\wh{\sf z}\btimes{\bf x}$. The toroidal projection of \eqref{eq:gy_primitive} yields the gyrokinetic toroidal angular-momentum equation \cite{Brizard_Tronko,Scott_Smirnov}
\begin{equation}
\pd{P_{\varphi}}{t} \;+\; \nabla\bdot{\bf Q}_{\varphi} \;=\; -\;\epsilon\;e\,\;\int F\;\pd{\langle\Phi_{1{\rm gc}}\rangle}{\varphi}\;d^{3}p,
\label{eq:P_varphi_dot}
\end{equation}
where
\[ P_{\varphi} \;\equiv\; {\bf P}\bdot\pd{\bf x}{\varphi} \;=\; \int\; F\,p_{{\rm gy}\varphi}\;d^{3}p \;\;\;{\rm and}\;\;\; {\bf Q}_{\varphi} \;\equiv\; \vb{\Pi}_{\rm gy}\bdot\pd{\bf x}{\varphi} \;=\; \int \;F\; \frac{d_{\rm gy}{\bf X}}{dt}\;p_{{\rm gy}\varphi} \;d^{3}p, \]
and we used the identity \eqref{eq:C_varphi} to obtain
\[ (\nabla\bdot\vb{\Pi}_{\rm gy})\bdot\pd{\bf x}{\varphi} \;=\; \nabla\bdot\left(\vb{\Pi}_{\rm gy}\bdot\pd{\bf x}{\varphi}\right) \;-\;
\vb{\Pi}_{\rm gy}^{\top}:\nabla\left(\pd{\bf x}{\varphi}\right) \;=\; \nabla\bdot{\bf Q}_{\varphi} \;+\; \int F\left( \frac{e}{c}\,
\pd{{\bf A}^{*}}{\varphi}\bdot\frac{d_{\rm gy}{\bf X}}{dt} \right). \]
Third, we introduce the operation of magnetic-surface average $\llbracket\;\cdots\;\rrbracket \equiv {\cal V}^{-1}\oint\;(\cdots)\;{\cal J}\;d\vartheta\,
d\varphi$, with the magnetic--coordinate $(\psi,\theta,\varphi)$ Jacobian ${\cal J} \equiv (\nabla\psi\btimes\nabla\theta\bdot\nabla\varphi)^{-1}$ and
${\cal V} \equiv \oint{\cal J}d\vartheta\,d\varphi$. Next, we introduce the gyrokinetic parallel-toroidal momentum
\begin{equation}
P_{\|\varphi} \;\equiv\; P_{\varphi} \;+\; \frac{\psi}{c}\;\varrho \;=\; \left( \int\,F\;p_{\|}\,d^{3}p\right)\,b_{\varphi},
\label{eq:P_par_phi}
\end{equation}
and obtain the surface-averaged gyrokinetic parallel-toroidal momentum equation
\begin{equation}
\pd{\llbracket P_{\|\varphi}\rrbracket}{t} \;=\; -\;\frac{1}{{\cal V}}\;\pd{}{\psi}\left( {\cal V}\frac{}{}\left\llbracket Q_{\varphi}^{\psi}\right\rrbracket\right) \;+\; \frac{\psi}{c}\;\pd{\llbracket\varrho\rrbracket}{t} \;-\; \epsilon\;e\,\left\llbracket\int\;F\;\pd{\langle\Phi_{1{\rm gc}}\rangle}{\varphi}\;d^{3}p \right\rrbracket,
\label{eq:P_par_phi_dot}
\end{equation}
where
\be
\llbracket Q_{\varphi}^{\psi}\rrbracket & = & \left\llbracket\int\;F\;\frac{d_{\rm gy}\psi}{dt}\;p_{{\rm gy}\varphi}\;d^{3}p\right\rrbracket \;\equiv\; \llbracket Q_{\|\varphi}^{\psi}\rrbracket \;-\; \frac{\psi}{c}\;\llbracket \nabla\psi\bdot{\bf J}\rrbracket,
\ee
and $\llbracket \nabla\psi\bdot{\bf J}\rrbracket \equiv e\;\llbracket\int\;F\;(d_{\rm gy}\psi/dt)\;d^{3}p\rrbracket$. Lastly, we use the gyrocenter charge conservation law
\begin{equation}
\pd{\llbracket\varrho\rrbracket}{t} \;=\; -\;\llbracket\nabla\bdot{\bf J}\rrbracket \;\equiv\; -\frac{1}{\cal V}\,\pd{}{\psi}\left({\cal V}\, \llbracket \nabla\psi\bdot{\bf J}\rrbracket \right),
\label{eq:gyro_charge}
\end{equation}
so that \eqref{eq:P_par_phi_dot} becomes
\begin{equation}
\pd{\llbracket P_{\|\varphi}\rrbracket}{t} \;=\; -\; \frac{1}{{\cal V}}\;\pd{}{\psi}\left( {\cal V}\;\left\llbracket Q_{\|\varphi}^{\psi}\right\rrbracket\right) \label{eq:gyro_tor_par} \;+\; e\left\llbracket\int F\left( \frac{1}{c}\frac{d_{\rm gy}\psi}{dt} -
\epsilon\pd{\langle\Phi_{1{\rm gc}}\rangle}{\varphi}\right)d^{3}p \right\rrbracket
\end{equation}
where $c^{-1}\,d_{\rm gy}\psi/dt - \epsilon\;\partial\langle\Phi_{1{\rm gc}}\rangle/\partial\varphi$ represents the gyrocenter toroidal electric field.
If we now use the gyrokinetic quasineutrality condition $\varrho \equiv \nabla\bdot\mathbb{P}$, the gyrocenter charge conservation law
\eqref{eq:gyro_charge} becomes (with ${\cal P}^{\psi} \equiv \nabla\psi\bdot\mathbb{P}$)
\[ \pd{\llbracket\varrho\rrbracket}{t} \;\equiv\; \left\llbracket\nabla\bdot\pd{\mathbb{P}}{t}\right\rrbracket \;=\; \frac{1}{{\cal V}}\;\pd{}{\psi}
\left( {\cal V}\;\pd{\llbracket{\cal P}^{\psi}\rrbracket}{t}\right) \]
which implies the gyrocenter ambipolarity condition $\llbracket J_{\rm phys}^{\psi}\rrbracket \equiv \llbracket(\partial{\cal P}^{\psi}/\partial t) + \nabla\psi\bdot{\bf J}\rrbracket \equiv 0$, where the magnetization-current contribution vanishes since $\llbracket\nabla\psi\bdot\nabla\btimes{\bf M}\rrbracket = \llbracket\nabla\bdot\left({\bf M}\btimes\nabla\psi\right)\rrbracket \equiv 0$. The parallel-toroidal momentum equation
\eqref{eq:gyro_tor_par} thus becomes
\begin{equation}
\pd{}{t}\left(\llbracket P_{\|\varphi}\rrbracket + \frac{1}{c}\;\llbracket{\cal P}^{\psi}\rrbracket \right) \;+\; \frac{1}{{\cal V}}\pd{}{\psi}
\left({\cal V}\;\llbracket Q_{\|\varphi}^{\psi}\rrbracket \right) \;=\; -\;\epsilon\;e\;\left\llbracket\int\;F\;
\pd{\langle\Phi_{1{\rm gc}}\rangle}{\varphi}\;d^{3}p \right\rrbracket.
\label{eq:total_toroidal}
\end{equation}
Here, the total toroidal-momentum density $\llbracket P_{\|\varphi}\rrbracket + c^{-1}\;\llbracket{\cal P}^{\psi}\rrbracket$ is derived from
the gyrocenter Vlasov moment of the toroidal gyrocenter velocity
\[ \pd{{\bf X}}{\varphi}\bdot m\,\frac{d_{\rm gy}{\bf X}}{dt} \;\equiv\; m\,R^{2}\;\frac{d_{\rm gy}\varphi}{dt} \;=\; p_{\|}\;b_{\varphi} \;-\;
\frac{\nabla\psi}{B\Omega}\bdot\left( \mu\,\nabla B \;+\; \frac{p_{\|}^{2}}{m}\;\bhat\bdot\nabla\bhat \;+\; \epsilon\,e\nabla\langle\Phi_{1{\rm gc}}\rangle\right), \]
where the first term contributes to $\llbracket P_{\|\varphi}\rrbracket$ while the second set of terms contribute to the radial gyrokinetic polarization
$\llbracket{\cal P}^{\psi}\rrbracket \equiv \llbracket{\cal P}_{\rm gc}^{\psi}\rrbracket + \epsilon\;\llbracket{\cal P}_{\rm gy}^{\psi}\rrbracket$.
As our last step, we perform a guiding-center multipole expansion: $\partial\langle\Phi_{1{\rm gc}}\rangle/\partial\varphi \equiv \partial\Phi_{1}/\partial\varphi + \langle\vb{\rho}_{\rm gc}\rangle\bdot\nabla(\partial\Phi_{1}/\partial\varphi) + \cdots$ so that we find
\[ e\,\int F\pd{\langle\Phi_{1{\rm gc}}\rangle}{\varphi}d^{3}p \simeq \left( \varrho_{\rm gy} -\frac{}{} \nabla\bdot\mathbb{P}_{\rm gc}\right)
\pd{\Phi_{1}}{\varphi} + \nabla\bdot\left( \mathbb{P}_{\rm gc}\pd{\Phi_{1}}{\varphi} \right) = \epsilon\left(\nabla\bdot\mathbb{P}_{\rm gy}\right)
\pd{\Phi_{1}}{\varphi} + \nabla\bdot\left( \mathbb{P}_{\rm gc}\pd{\Phi_{1}}{\varphi} \right), \]
where we used the gyrokinetic quasineutrality condition $\varrho_{\rm gy} - \nabla\bdot\mathbb{P}_{\rm gc} = \epsilon\,\nabla\bdot\mathbb{P}_{\rm gy}$.
Hence, the parallel-toroidal momentum equation \eqref{eq:total_toroidal} becomes
\begin{equation}
\pd{}{t}\left\llbracket P_{\|\varphi} \;+\; \frac{1}{c}\;{\cal P}^{\psi} \right\rrbracket \;=\; -\;\frac{1}{{\cal V}}\pd{}{\psi}\left({\cal V}\frac{}{}\left\llbracket Q_{\|\varphi}^{\psi} \;+\; \epsilon\,{\cal P}_{\rm gc}^{\psi}\;\pd{\Phi_{1}}{\varphi} \right\rrbracket \right) \;-\;\epsilon^{2}\;\left\llbracket\left(\nabla\bdot\mathbb{P}_{\rm gy}\right)\;\pd{\Phi_{1}}{\varphi}\right\rrbracket.
\label{eq:total_tor_mom}
\end{equation}
We note that, in the zero-Larmor-radius approximation $\mathbb{P}_{\rm gy} \simeq (mnc^{2}/B^{2})\;\nabla_{\bot}\Phi_{1}$, we find
\[ \left\llbracket\left(\nabla\bdot\mathbb{P}_{\rm gy}\right)\;\pd{\Phi_{1}}{\varphi}\right\rrbracket \simeq \left\llbracket\nabla\bdot\left(
\mathbb{P}_{\rm gy}\pd{\Phi_{1}}{\varphi}\right) - \pd{}{\varphi}\left( \frac{mnc^{2}}{2\;B^{2}}|\nabla_{\bot}\Phi_{1}|^{2} \right)\right\rrbracket = \frac{1}{\cal V}\pd{}{\psi}\left({\cal V}\;\left\llbracket {\cal P}_{\rm gy}^{\psi}\pd{\Phi_{1}}{\varphi}\right\rrbracket\right), \]
so that \eqref{eq:total_tor_mom} becomes the gyrokinetic toroidal angular-momentum conservation law for the truncated gyrokinetic Vlasov-Poisson equations
(compare with equation 98 of \cite{Scott_Smirnov})
\begin{equation}
\pd{}{t}\left\llbracket P_{\|\varphi} \;+\; \frac{1}{c}\;{\cal P}^{\psi} \right\rrbracket \;=\; -\;\frac{1}{{\cal V}}\pd{}{\psi}\left({\cal V}\frac{}{}\left\llbracket Q_{\|\varphi}^{\psi} \;+\; \epsilon\,{\cal P}^{\psi}\;\pd{\Phi_{1}}{\varphi} \right\rrbracket \right),
\end{equation}
which includes guiding-center and gyrocenter polarization effects.

Work by AJB was supported by a U.~S.~DoE grant under contract No.~DE-FG02-09ER55005. NT was supported by the Engineering and Physical Sciences Research Council grant $\mathrm{EP/H049460/1}$ for the EPS conference participation.

\end{document}